\documentclass[journal=jctcce,manuscript=article,layout=twocolumn]{achemso}
\usepackage[T1]{fontenc}       % Use modern font encodings
\usepackage[english]{babel}
\usepackage{amsmath,amssymb}
\usepackage{graphicx,color}

\usepackage[fontsize=11pt]{scrextend}
\def\QE{\textsc{Quantum ESPRESSO}}
\definecolor{tangerine}{rgb}{0.944,0.522,0}

\title{Multimodel Approach to the Optical Properties of Molecular
  Dyes in Solution}
\author{Iurii Timrov}
\affiliation[SISSA]{SISSA -- Scuola Internazionale Superiore di Studi
  Avanzati\\ Via Bonomea 265, 34136 Trieste, Italy}
\altaffiliation{Current address: Theory and Simulation of Materials
  (THEOS), \'Ecole Polytechnique F\'ed\'erale de Lausanne, 1015 Lausanne,
 Switzerland} 
\author{Marco Micciarelli}
\author{Marta Rosa}
\affiliation[SISSA]{SISSA -- Scuola Internazionale Superiore di Studi
  Avanzati\\ Via Bonomea 265, 34136 Trieste, Italy} 
\author{Arrigo Calzolari}
\affiliation[CNR-NANO]{CNR-NANO, Istituto Nanoscienze, Centro S3\\ Via
  Campi 213A, 41125 Modena, Italy} 
\author{Stefano Baroni}
\email{baroni@sissa.it}
\affiliation[SISSA]{SISSA -- Scuola Internazionale Superiore di Studi
  Avanzati\\ Via Bonomea 265, 34136 Trieste, Italy} 

\begin{document}

\begin{abstract}
  We introduce a multimodel approach to the simulation of the optical
  properties of molecular dyes in solution, whereby the effects of
  thermal fluctuations and of dielectric screening on the absorption
  spectra are accounted for by explicit and implicit solvation models,
  respectively. Thermal effects are treated by averaging the spectra
  of molecular configurations generated by an \emph{ab initio}
  molecular-dynamics simulation where solvent molecules are treated
  explicitly. Dielectric effects are then dealt with implicitly by
  computing the spectra upon removal of the solvent molecules and
  their replacement with an effective medium, in the spirit of a
  continuum solvation model. Our multimodel approach is validated by
  comparing its predictions with those of a fully explicit-solvation
  simulation for cyanidin-3-glucoside (cyanin) chromophore in
  water. While multimodel and fully explicit-solvent spectra may
  differ considerably for individual configurations along the
  trajectory, their time averages are remarkably similar, thus
  providing a solid benchmark of the former, and allowing us to save
  considerably on the computer resources needed to predict accurate
  absorption spectra. The power of the proposed methodology is finally
  demonstrated by the excellent agreement between its predictions and
  the absorption spectra of cyanin measured at strong and intermediate
  acidity conditions.
\end{abstract}

\section{Introduction}
\label{sec:introduction}

Modeling the optical properties of molecular dyes in solution has long
relied on a static picture, whereby thermal fluctuations are
thoroughly ignored, either by neglecting the effects of the
environment altogether, or by accounting for the dielectric screening
of the solvent through some kind of {\it continuum solvation model}
(CSM),\cite{Tomasi:2005} such as the {\it polarizable continuum model}
(PCM),\cite{Miertus:1981, Fortunelli:1994, Mennucci:1997, Barone:1997,
  Mennucci:2010} the {\it conductor-like screening model}
(COSMO),\cite{Klamt:1993, Stefanovich:1994b, Truong:1995, Barone:1998}
or the (revised) {\it self-consistent continuum solvation} (SCCS)
model.\cite{Fattebert:2002, Fattebert:2003, Scherlis:2006,
  Dziedzic:2011, Andreussi:2012, Dupont:2013, Andreussi:2014} In both
cases, the optical spectra are often obtained at the molecular
minimum-energy configuration, as computed \emph{in vacuo} or in some
effective medium. Nevertheless, recent studies have shown the
importance of thermal fluctuations on the optical properties of
solvated dyes.
\cite{Barone:2010,Malcioglu:2011b,Jacquemin:2011,DeMitri:2013,Ge:2015b,Cacelli:2016}
Thermal fluctuations are accurately described in \emph{molecular
  solvation models} (MSMs), in which solvent molecules are treated
explicitly, \emph{i.e.} on the same atomistic ground as the solute
(this approach is often referred to as \emph{explicit solvent}
model). Some of the previous attempts to account for thermal
  fluctuations in the optical properties of solvated molecules were
  based on molecular dynamics (MD) utilizing classical force fields
  accurately targeted to the specific system(s) of
  interest. \cite{Barone:2010,DeMitri:2013,Cacelli:2016} While this
  approach is apt to cope with thermal effects on the optical
  absorption spectra, its practical implementation is hampered by the
  limited transferability of the available force fields, which would
  require extensive fitting and validation procedures for each system
  to be examined. In one of our previous works\cite{Malcioglu:2011b}
we introduced an MSM approach to simulating the optical properties of
solvated dyes where the entire system (solute plus solvent) is treated
quantum mechanically and let evolve according to \emph{ab initio}
(AI) molecular dynamics,\cite{Car:1985, Marx:2009} while
absorption spectra are computed on the fly using time-dependent
density-functional theory (TDDFT),\cite{Runge:1984, Marques:2012} and
averaged along the trajectory. This naturally accounts for both
thermal and dielectric solvation effects on the optical spectra,
  yet not requiring any unwieldy system-specific fitting or
  validation. When it comes to simulating large systems comprising
hundreds of atoms, however, the deployment of \emph{ab initio}
MSMs is hindered by the large computational cost of hybrid
functionals,\cite{Becke:1993,Stephens:1994,Perdew:1996b,Adamo:1999}
particularly when using plane-wave basis sets, as it is common
practice in many applications involving AIMD. In a previous
  paper\cite{Ge:2015b} we proposed to use a GGA
  exchange-correlation (XC) energy functional\cite{Perdew:1996} to
  generate explicit-solvent AIMD trajectories and a semi-empirical
  \emph{morphing procedure}, tailored on the gas-phase
  B3LYP\cite{Stephens:1994} TDDFT spectra of the dye, to correct the
  GGA spectra evaluated along the the AIMD trajectory. GGA is
  generally assumed to be accurate enough for the sake of sampling
  AIMD trajectories, in spite of the minor, but not always negligible,
  differences existing between the harmonic frequencies computed using
  different functionals.\cite{Zhao:2006} As for the morphing procedure
  previously proposed, while it proved accurate enough to deal with
  global properties of the optical spectra, such as the color
  expressed by a given solution, it may not be so when it comes to
  predicting individual spectral features.

In this paper we propose an approach to this problem that combines the
advantages of CSMs and MSMs, by applying them selectively to different
steps of a multimodel simulation protocol, while avoiding or
mitigating some of their drawbacks, such as the neglect of thermal
fluctuations in CSMs and the high computational cost of MSMs, and
  without using any semi-empirical procedure. In our
approach thermal fluctuations are accounted for by using AIMD in
conjunction with MSM for the solvent using GGA energy
  functional. Optical spectra are then computed on the fly using
TDDFT and a hybrid functional such as B3LYP,\cite{Stephens:1994} which
is well suited for spectroscopic purposes,\cite{Charaf-Eddin:2013} and
averaged along the AIMD trajectory. In order to save on computer
resources, this step is performed by representing solvent molecules
with an effective medium.\cite{Timrov:2015}

\begin{figure}[t]
  \begin{center}
    \includegraphics[width=0.31\textwidth]{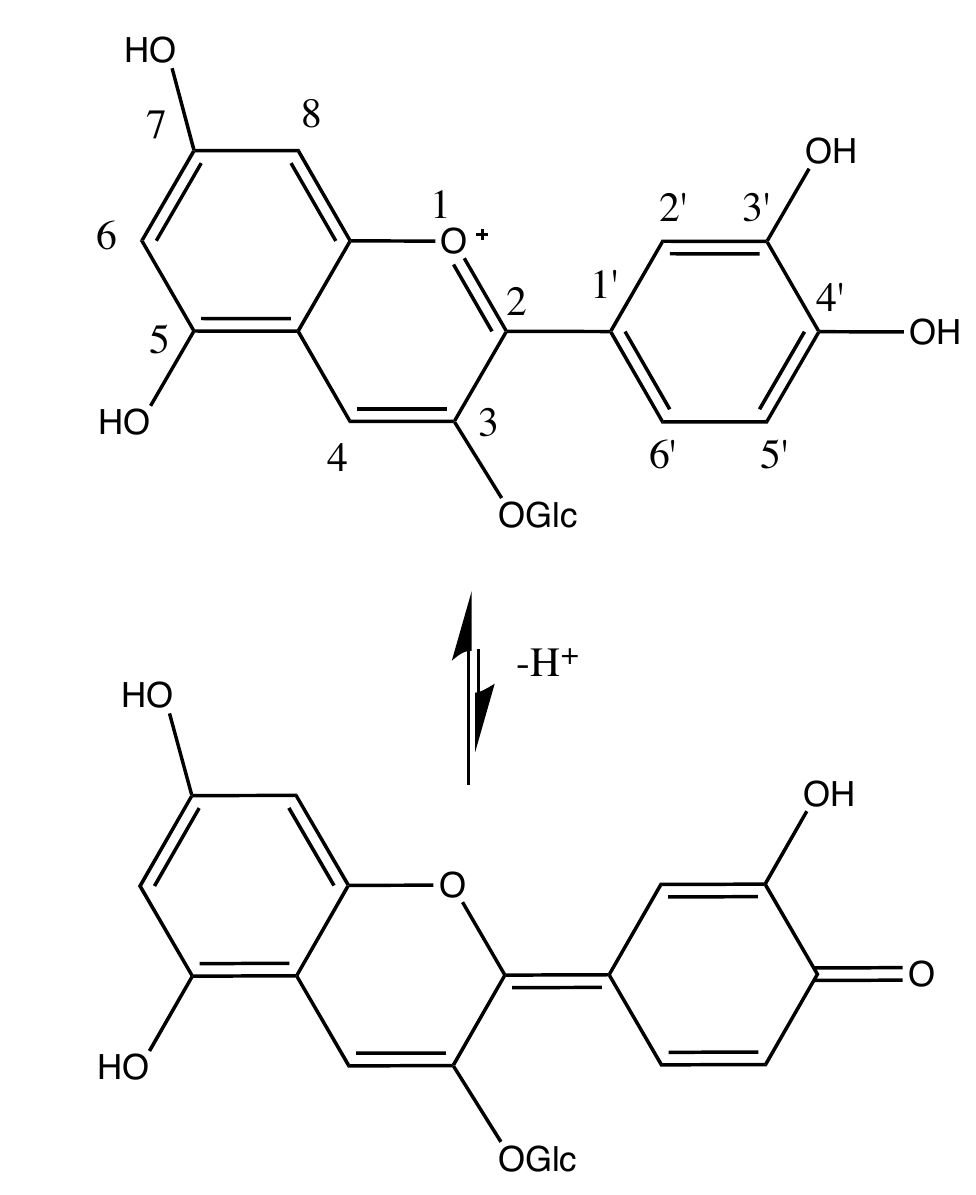}
  \end{center}
  \caption{Chemical structure of the cyanin flavylium cation AH$^+$
    (top) and quinonoidal base A (bottom). The ``Glc'' label indicates
    the attached glucose group. The quinonoidal base differs from
      the flavylium cation by the deprotonation of the hydroxyl group
      attached at position $4'$.}
  \label{fig:cyanin}
\end{figure}

The combination of MSM and CSM methods into the same simulation
protocol is the cornerstone of our approach: while CSMs have been
extensively tested for individual molecular configurations
(\emph{e.g.} the minimum-energy geometry), the effects of CSMs on
configurational averages of the solute spectra have not been addressed
before. Hence, our approach is thoroughly tested by comparing its
predictions with those of a pure MSM in the paradigmatic case of
cyanidin-3-glucoside (cyanin) in its neutral charge state (quinonoidal
base, A), which is stable at weakly acidic conditions (see
Figure~\ref{fig:cyanin}). \cite{Brouillard:1988,Gould:2009} Cyanin
features both a $\pi$-aromatic skeleton and lateral hydroxyl
terminations, thus being representative of different out-of-plane
(hydrophobic) and in-plane (hydrophilic) solute-solvent
interactions.\cite{Calzolari:2010} This comparison, which results in a
high level of agreement, provides a solid benchmark of CSMs and
demonstrate that, while spectra obtained from CSMs and MSMs may differ
considerably for individual configurations along the AIMD trajectory,
these differences tend to vanish upon time averaging. The
power of our methodology is further and significantly confirmed by
comparing its predictions with experiment both at strongly acidic
conditions, where the positive flavylium cation (AH$^+$) is stable,
and at weakly acidic conditions, where the flavylium cation and the
quinonoidal base (A) coexist (see
Figure~\ref{fig:cyanin}). \cite{Brouillard:1988,Gould:2009}

The rest of the paper is organized as follows. In
Sec.~\ref{sec:method} we introduce our simulation protocol; in
Sec.~\ref{sec:validation} this protocol is validated by a detailed
comparison of its predictions for the quinonoidal base A with those of
a fully explicit MSM; in Sec.~\ref{sec:expt} the predictive power of
our multimodel approach is demonstrated against experimental
absorption spectra, measured at strongly and weakly acidic conditions;
Sec.~\ref{sec:conclusions} finally contains our conclusions.

\begin{figure*} [t]
  \begin{center}
    \hbox to \hsize{
    \hfil\includegraphics[width=0.7\columnwidth]{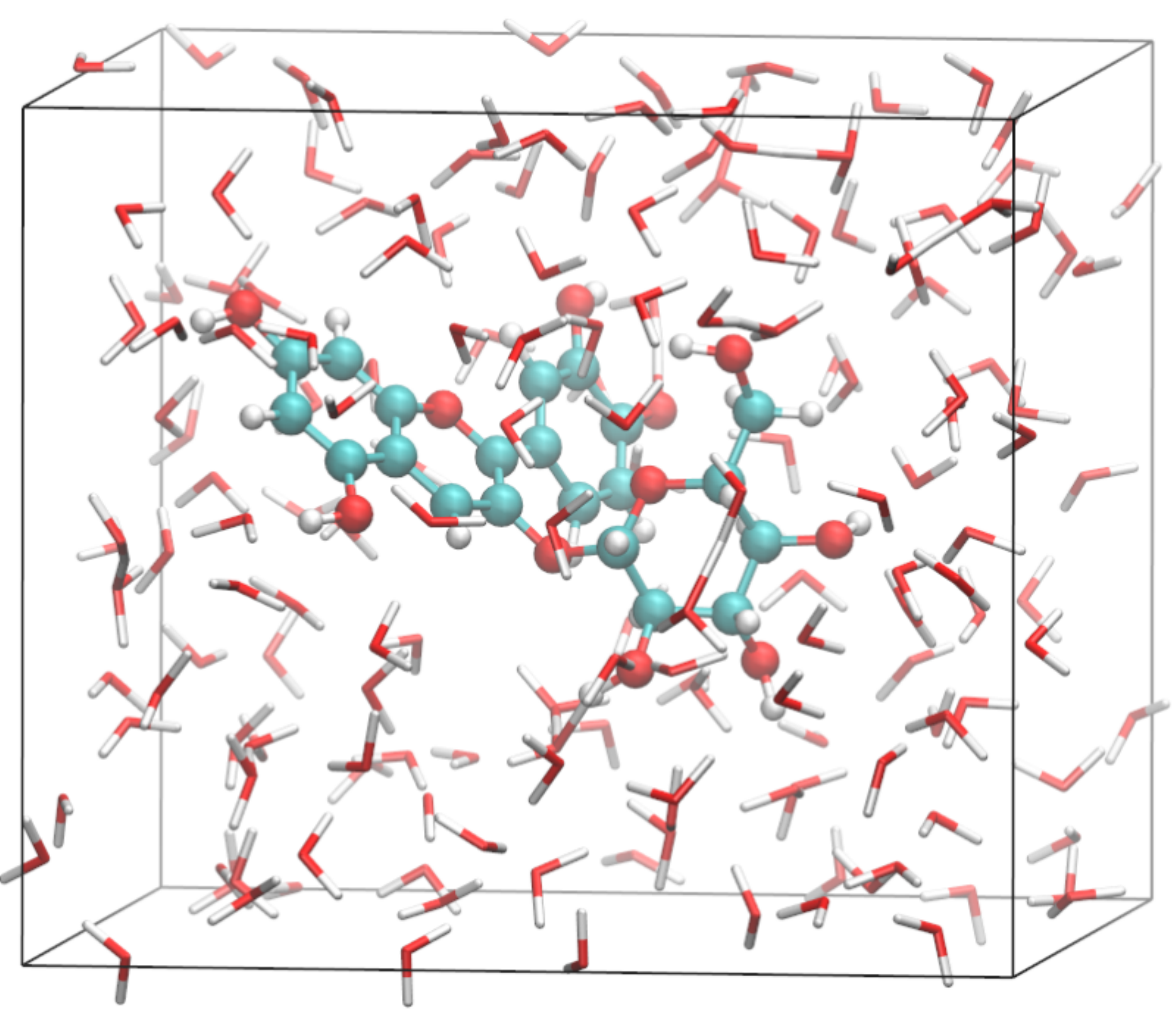}
    \hfil\includegraphics[width=0.7\columnwidth]{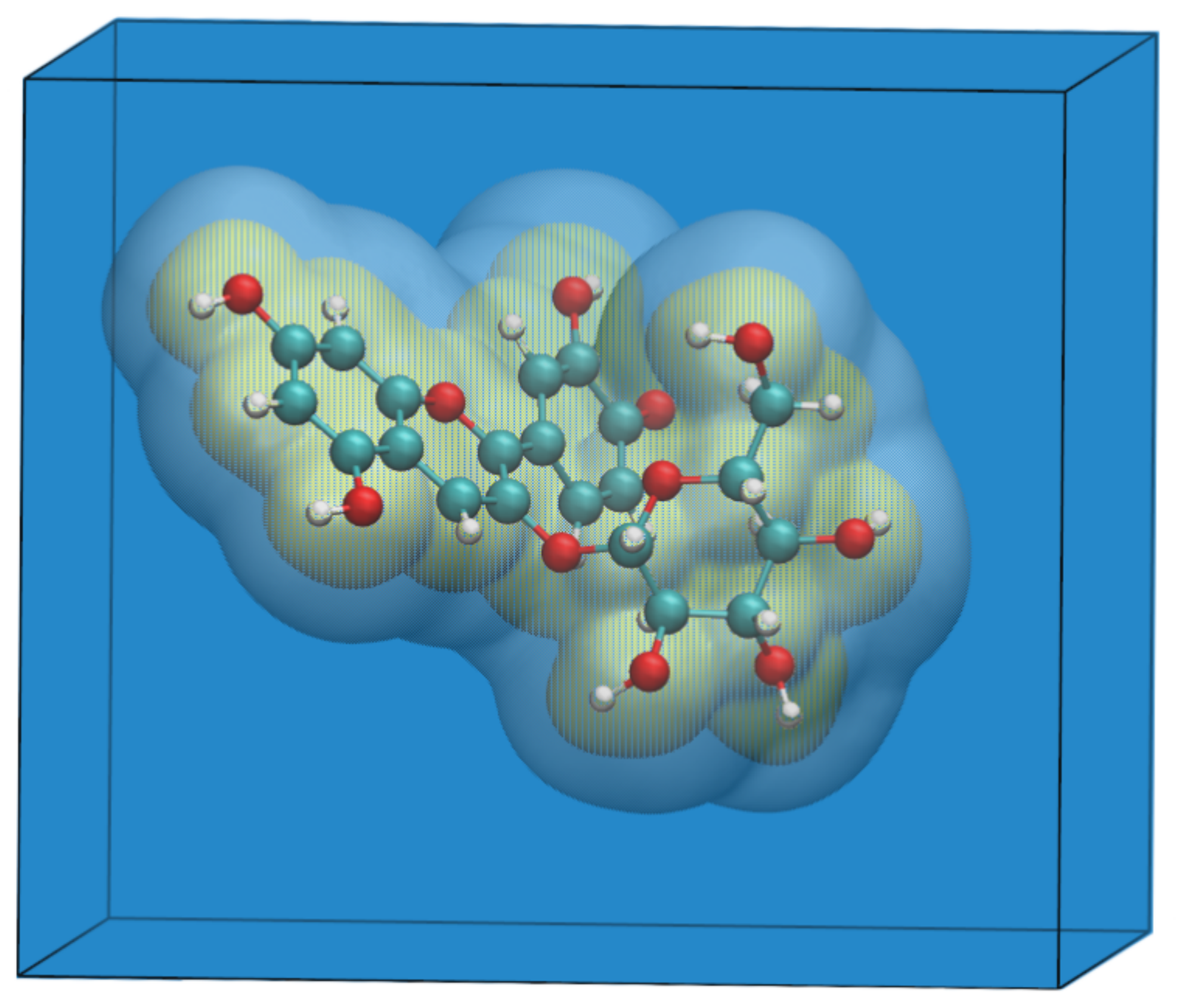}
    \hfil
    }
    \caption{Left panel: Explicit solvent model of water-solvated
      cyanin. A cyanin molecule is surrounded by a number of water
      molecules, which are treated fully quantum mechanically in a
      periodically repeated supercell. Right panel: The SCCS model for
      neutral cyanin in solution. The cavity is defined by two
      surfaces, corresponding to two different iso-values of the
      electron density, and represented in light green and light
      blue. Inside the internal surface (light green), the dielectric
      constant of the medium is 1; outside the external one (light
      blue) it equals the value of the solvent; whereas it is
      intermediate in-between. The color code for the atoms is:
      H--white, O--red, C--cyan.}
    \label{fig:2models}
  \end{center}
\end{figure*}

\section{Methodology} \label{sec:method}
Our multimodel approach consists of the following two-step protocol:

\noindent\textbf{MSM-AIMD --} We first perform a
tens-of-picoseconds-long AIMD simulation of the dye in solution at
ambient conditions, where the solvent is treated explicitly at the MSM
level (see Figure~\ref{fig:2models}, left panel).  In view of the
minor effects of hybrid functionals on the magnitude of thermal
fluctations, and of their considerable computational cost, this step
can be often conveniently performed at the GGA level of theory. A
number of statistically independent molecular configurations
(\emph{frames}) is selected along the trajectory, typically a fraction
of ps apart one from the other.

\noindent\textbf{CSM-TDDFT --} For each frame thus generated, we then
compute the TDDFT spectrum of the system obtained by stripping off the
water molecules that surround the solute, and mimicking their effects
with a continuum solvation model, such as the SCCS model (see
Figure~\ref{fig:2models}, right panel). Due to the poor performance of
GGA in computing excited state properties, this step is best performed
using a hybrid XC functional such as
B3LYP.\cite{Stephens:1994,Charaf-Eddin:2013} The spectra thus computed
for each individual frame are finally time-averaged over the entire
trajectory.

%\noindent\textbf{IMPLEMENTATION --}
\subsection{Implementation}
The multimodel simulation protocol introduced above has been
implemented using the \QE\ suite of computer
codes,\cite{Giannozzi:2009} which provides specific modules to perform
standard density functional theory (DFT) calculations (\texttt{pw.x}),
AIMD simulations (\texttt{cp.x}), and to compute optical spectra
within time-dependent density functional {\it perturbation} theory
(TDDFpT: \texttt{turboTDDFT}).\cite{Walker:2007, Rocca:2008,
  Baroni:2010, Baroni:2012c, Malcioglu:2011, Ge:2014} The SCCS model
used to simulate the effects of dielectric screening on the structure
of solvated molecules is implemented in the \textsc{Environ} add-on to
\QE,\cite{Andreussi:2012} recently extended to
TDDFpT.\cite{Timrov:2015}

All calculations were performed using the plane-wave (PW)
pseudopotential (PP) method, with a tetragonal supercell of dimensions
$19.2 \times 19.2 \times 16.0$~\AA$^3$. These dimensions were chosen
to accommodate 171 H$_2$O molecules, mimicking the first solvation
shell at ambient conditions (Figure~\ref{fig:2models}, left panel).
The resulting system consisted of 565 atoms and 1538 valence electrons
for neutral cyanin.  In the case of flavylium cation the number of
electrons is the same, while the structure counts one extra proton
(Figure \ref{fig:cyanin}). Electrostatic divergences have been removed
in this case by introducing a neutralizing uniform background.  This
simulation setup is adequate to model the effect of the solvent, while
keeping the interaction of cyanin with its fictitious periodic images
small enough as to be negligible.

We have used the Perdew-Burke-Ernzerhof (PBE)\cite{Perdew:1996} GGA XC
functional, when doing AIMD simulations, while both PBE and the
B3LYP\cite{Stephens:1994} functionals have been used for computing
optical spectra, for benchmarking and production purposes,
respectively.  Ultra-soft and norm-conserving PPs from the \QE\ public
repository\cite{Timrov:Note:2016:PP} were were used in conjunction
with the PBE and B3LYP functionals, respectively.  Kohn-Sham orbitals
have been expanded in PWs up to a kinetic-energy cutoff of 25 and
50~Ry when using ultra-soft and norm-conserving PPs, respectively, and
in both cases the electronic charge-density and potentials have been
expanded in PWs up to 200~Ry. A reduced cutoff of 50~Ry was used to
implement the Fock-exchange operator in the B3LYP
functional. \cite{Ge:2014}

In the SCCS model the homogeneous dielectric medium is characterized
by the experimental dielectric constant of the solvent, which in the
case of water at $T=298$~K equals 78.5 and 1.8 in the static and
optical regimes, respectively.\cite{Mennucci:1999, Cammi:1999} The
solute is hosted by a cavity defined by the molecular ground-state
electron charge-density distribution, $n(\mathbf{r})$, such that the
dielectric constant goes smoothly from 1 inside (when
$n> 5 \times 10^{-3}~\mathrm{a.u.}$) to the value appropriate to the
solvent outside (when $n< 10^{-4}~\mathrm{a.u.}$).
\cite{Andreussi:2012, Timrov:2015} The resulting scheme is illustrated
in Figure~\ref{fig:2models}~(right panel). Only electrostatic
screening is considered in our protocol, while all non-electrostatic
effects (such as Pauli repulsion, dispersion, cavitation, etc.) are
consistently disregarded.

\begin{figure}[t]
  \begin{center}
    \includegraphics[width=0.38\textwidth]{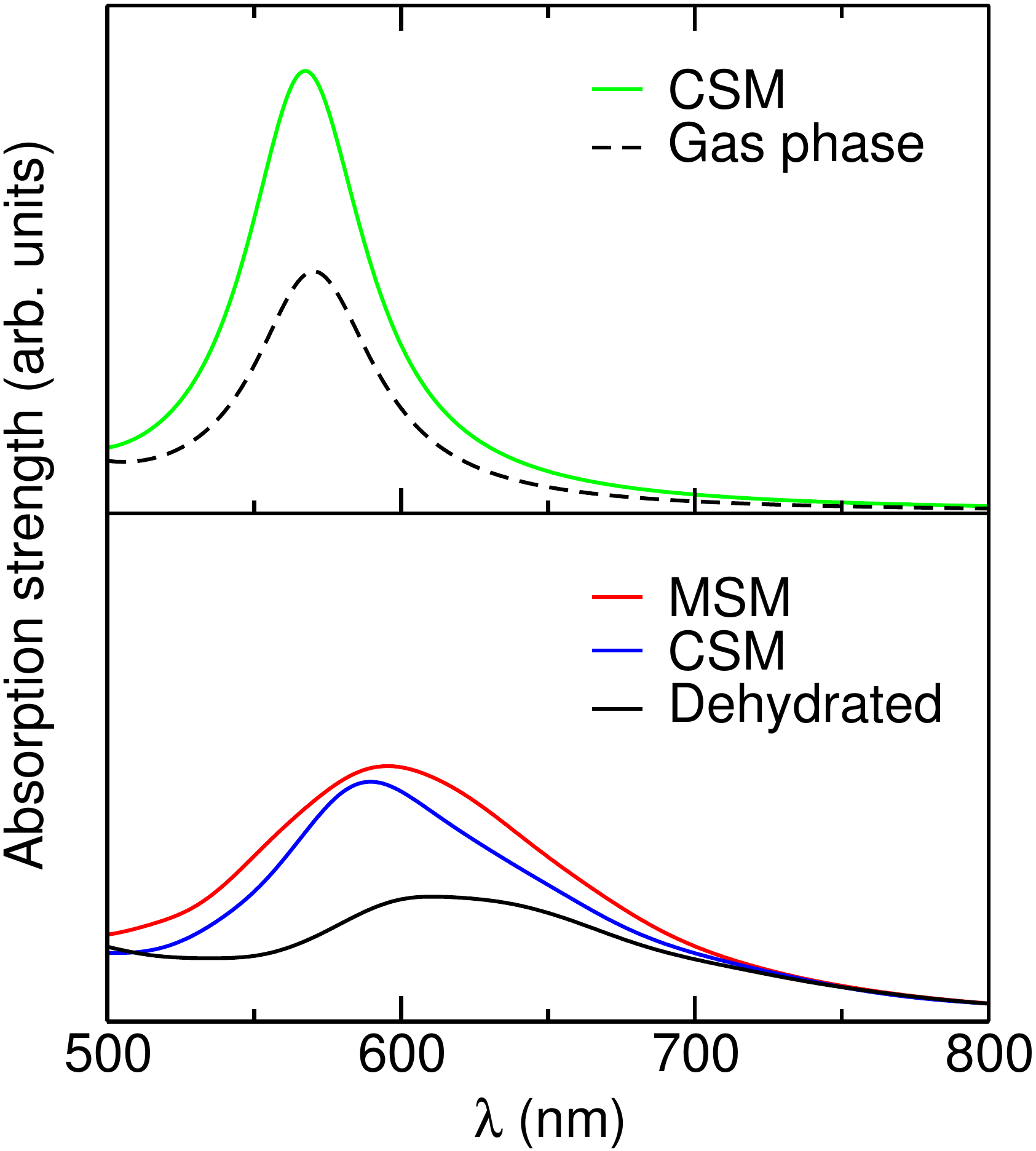}
  \end{center}
  \caption{Spectra obtained with the GGA-PBE XC functional. Upper
    panel: Optical absorption spectra of the neutral cyanin molecule
    in its minimum-energy configuration in the gas phase and in the
    continuum solvent. Lower panel: Time-averaged optical absorption
    spectra computed for the dehydrated molecular structure, and for
    the molecular structure including either the molecular or
    continuum solvent Relative intensities on all figures are
    respected.}
  \label{fig:comparison_PBE}
\end{figure}

AIMD was performed using the Car-Parrinello extended Lagrangian
formalism,\cite{Car:1985, Pastore:1991} with a fictitious electronic
mass $\mu=\mathrm{600~a.u.}$ and a time step
$\Delta t= \mathrm{0.12~fs}$.  In order to facilitate the approach to
equilibrium and enhance dynamical sampling, all the atomic masses were
set to a same value of $\mathrm{M=12~a.m.u.}$ Dispersion forces have
been accounted for by using the semi-empirical Grimme's DFT-D2 van
der-Waals interaction correction.\cite{Grimme:2006, Barone:2009} Our
AIMD run consisted of an equilibration phase of $\sim 2.5~\mathrm{ps}$
at $300$~K using a Nos\'e-Hoover thermostat,\cite{Nose:1984b,
  Nose:1984c, Hoover:1985} followed by a 50~ps production trajectory.
The AIMD trajectory thus generated was sampled every
$\mathrm{0.5~ps}$,\footnote{We estimate that the
      auto-correlation time of the HOMO-LUMO gap is smaller than 100
      fs, while the typical auto-correlation times of solvent-solute
      hydrogen bonds is of the order of the ps.} resulting in 100
statistically independent frames for which the absorption spectra have
been computed by either keeping water molecules at their instantaneous
positions (MSM), or by representing them with an effective continuous
medium (CSM). The calculation of the absorption spectrum at every
frame has been performed in two steps: \emph{i)} we have first
computed the ground-state electronic structure, using the
\texttt{pw.x} code; \emph{ii)}~then, we have computed excitation
energies and oscillator strengths by solving Casida's
equation\cite{Casida:1996} with a Davidson-like
algorithm\cite{Tretiak:2009, Ge:2014} implemented in the latest
release of the \texttt{turboTDDFT} code.\cite{Malcioglu:2011, Ge:2014}
Absorption spectra have been convoluted with a Lorentzian broadening
function with a full width at half maximum of $\approx 0.1$~eV.  The
RGB representation of the color expressed by solvated dyes has been
obtained by using the {\em tristimulus} colorimetry theory, as
described \emph{e.g.} in Refs.~\citenum{Malcioglu:2011b} and
\citenum{Ge:2015b}.

For the sake of comparison, we have also calculated the absorption
spectra of the molecule \emph{in vacuo}, at the various geometries
generated by the AIMD trajectory upon removal of the water molecules
(we refer to the resulting spectra as the {\it dehydrated} spectra).
We stress that no structure relaxation has been performed for these
structures, which were thus kept fixed at the instantaneous geometries
generated by AIMD, so as to account for the effects of thermal
vibrations.

\section{Validating the multimodel approach}\label{sec:validation}

\begin{figure*}[t]
  \begin{center}
    \includegraphics[width=0.8\textwidth]{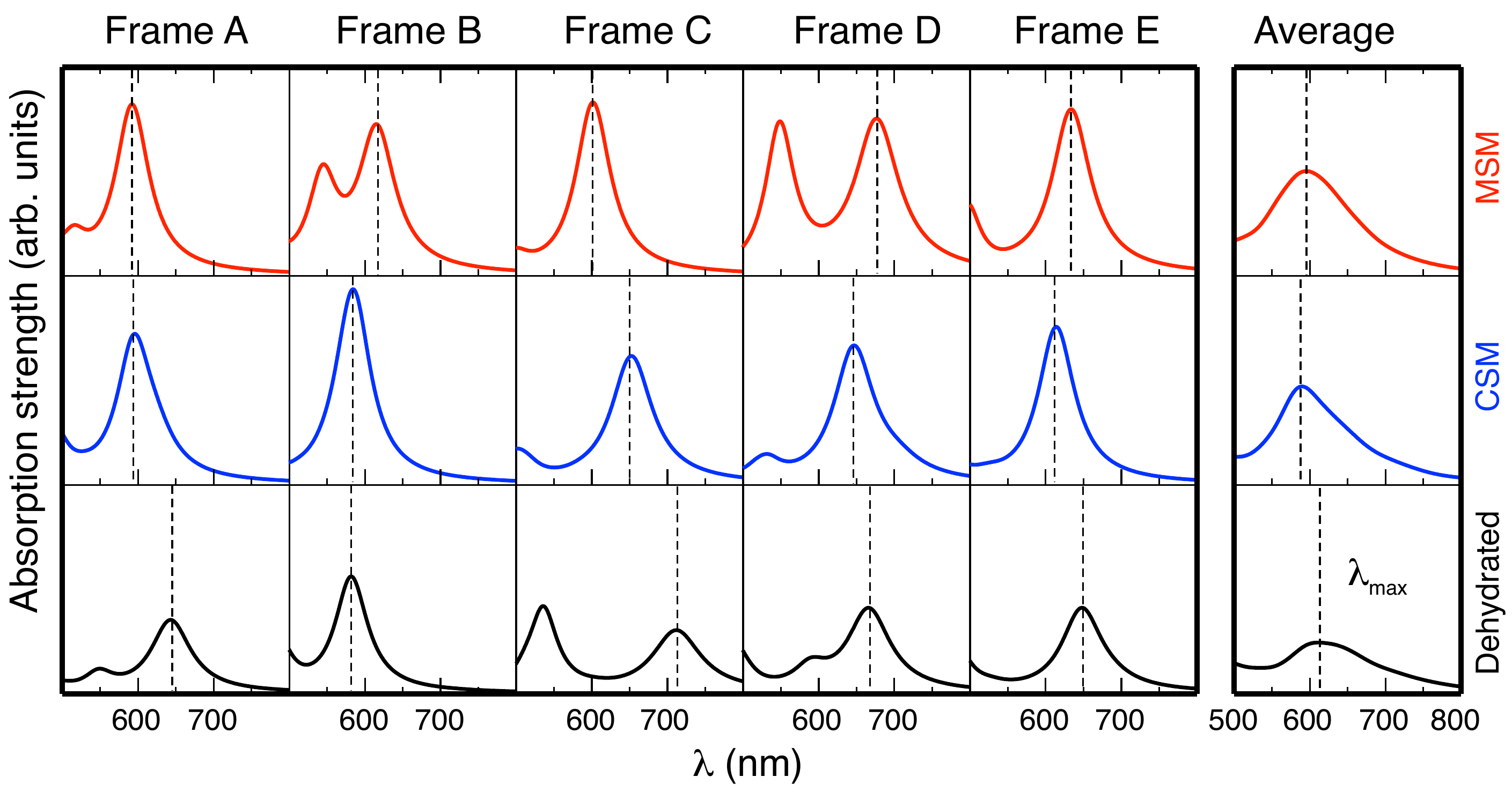}
  \end{center}
  \caption{Optical absorption spectra of the water-solvated neutral
    cyanin computed with the GGA-PBE XC functional using the molecular
    solvent (upper row), continuum solvent (central row), 
    and the dehydrated case (bottom row). First five columns represent optical
    absorption spectra for selected frames from the AIMD trajectory
    (labeled from A to E), while the last column represents the
    time-averaged spectra over 100 frames. Vertical dashed lines indicate positions of
    the first intense absorption peak at $\lambda_\mathrm{max}$. }
  \label{fig:many_spectra}
\end{figure*}

In this section the quality of our multimodel approach is assessed
against the predictions of a full MSM simulation of the absorption
spectra. For the sake of validation, we consider the case of the quinonoidal
base A.  The comparison is done at the same level of theory. The size
of the explicit model used here (171 water molecules, see above) would
make B3LYP calculations extremely expensive. On the other hand, the
computation of absorption spectra for this model can be afforded using
the PBE functional, which, although inadequate for quantitative
predictions, can indeed capture the qualitative features of the
optical spectra of anthocyanins.\cite{Ge:2015} In fact, while GGA XC
functionals underestimate the HOMO-LUMO gap and other low-lying
transition energies, the relative order of the the first few of them,
as well as the character of the molecular orbitals participating in
them are correctly predicted. For this reason our assessment was
performed using the PBE functional, while a B3LYP functional will be
used for the comparison with experiments (see Sec.~\ref{sec:expt}),
for which no full MSM simulation was attempted.
 
We start by comparing the lowest energy (i.e. longest-wavelength)
transitions of the molecule in the gas phase, as computed at the
minimum-energy geometry configuration {\em in vacuo}, with that
resulting from CSM, as computed at the SCCS minimum-energy
configuration. The two spectra (Figure~\ref{fig:comparison_PBE}, upper
panel) are characterized by an intense peak at
$\lambda_\mathrm{max} \approx 570~\mathrm{nm}$.\cite{Timrov:2015}
Accounting for dielectric screening results in a blue shift of
the first absorption peak by a few nanometers only and to an increase
of its intensity.

In order to evaluate the impact of thermal fluctuations independently
of solvent polarization effects, in
Figure~\ref{fig:comparison_PBE}~(lower panel) we report the
\emph{dehydrated} spectrum (\emph{i.e.} obtained by averaging the
spectra computed {\em in vacuo} for the molecular geometries sampled
by the AIMD trajectory), featuring a large broadening and a red shift
of the first absorption peak by
$\Delta\lambda_\mathrm{max} \approx 50~\mathrm{nm}$.  An explicit
account of both dielectric-screening and thermal effects is achieved
in both the MSM and multimodel approaches, as illustrated in Figure~
\ref{fig:comparison_PBE}, resulting in a markedly different behavior
from that obtained by neglecting either one of these effects.  The
time-averaged spectra from MSM and multimodel simulations agree
remarkably well, with minor residual differences consisting in a
deviation in $\lambda_\mathrm{max}$ of a few nm, and in a more
pronounced broadening predicted by MSM. The latter effect is likely
due to the larger fluctuations of the electrostatic potential induced
by the molecular nature of the solvent and/or to the neglect of
non-electrostatic solute-solvent interactions in our implementation of
the SCCS model.

\begin{figure}[t]
  \begin{center}
    \includegraphics[width=0.47\textwidth]{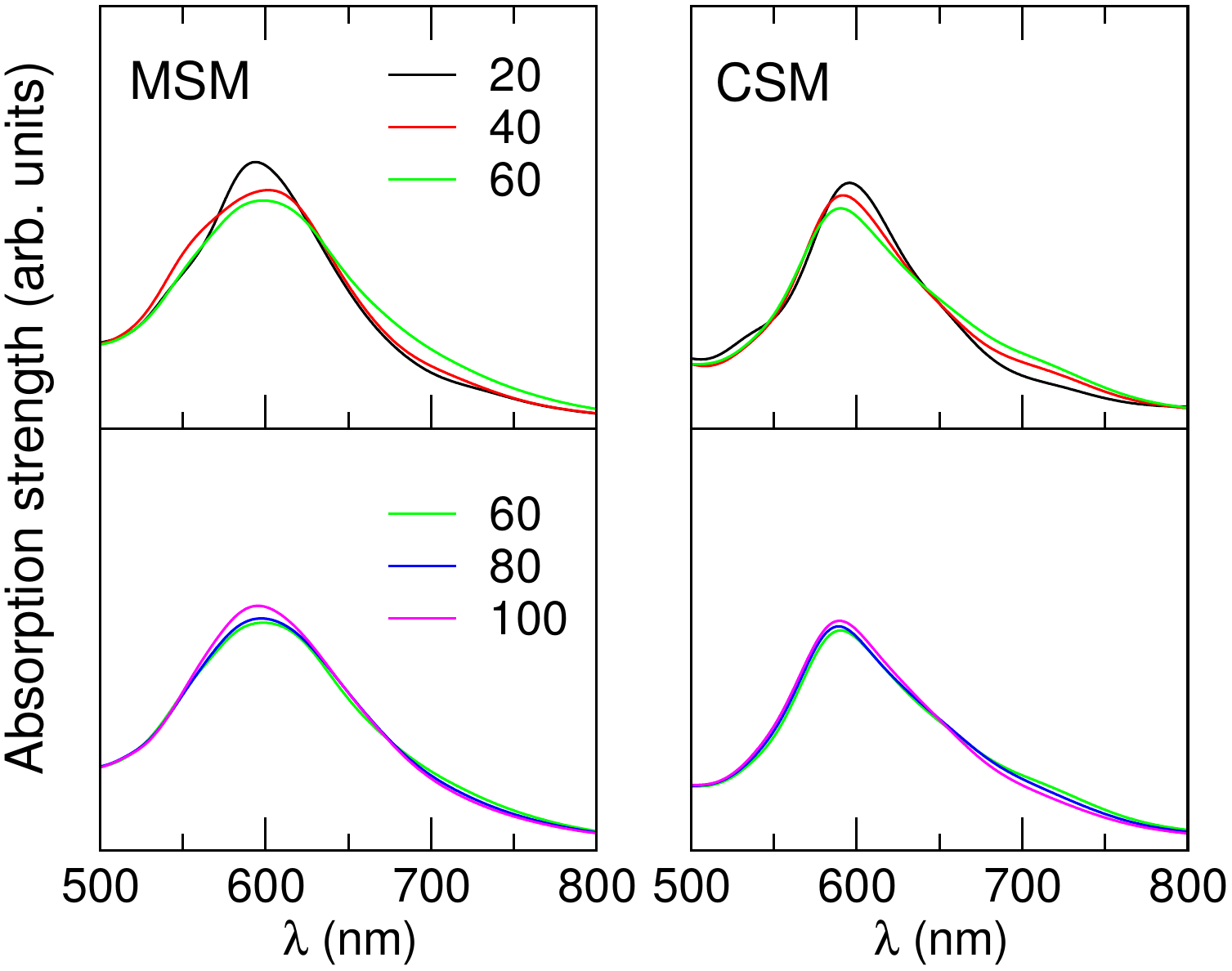}
  \end{center}
  \caption{Convergence of the time-averaged optical absorption
    spectrum of the neutral cyanin with respect to the number of
    frames from AIMD, using the GGA-PBE XC functional in the molecular
    (left) and the continuum (right) solvation model.  Top panels
    represent the time-averaged spectra over 20, 40, and 60 frames,
    while the bottom panels represent the time-averaged spectra over
    60, 80, and 100 frames.}
  \label{fig:convergence}
\end{figure}

Figure~\ref{fig:many_spectra} reports the MSM, CSM, and dehydrated
spectra computed at a few individual and non-consecutive frames,
representative of the full AIMD trajectory. Comparing the spectra
obtained at different levels of approximation for different frames
does not reveal a consistent picture. Sometimes different
approximations result in very similar spectra (Frame~A); in the
majority of cases, however, the value of $\lambda_\mathrm{max}$
predicted by different solvation models do not coincide (Frames~B --
E) and, moreover, the shape of the spectrum can differ considerably,
featuring one or two intense peaks in the range 500 -- 800 nm
(Frames~B and D). The difference between MSM and CSM spectra for a
same configuration of the solute is obviously a manifestation of the
discrete (molecular) nature of the solvent. In the present case,
inspection of the microscopic configurations of water molecules around
the solute reveals a noticeable correlation between the number of
hydrogen bonds formed by water molecules and the most acidic O atom in
position 4' (see Fig. 1) and the difference between the MSM and CSM
wavelengths of the fundamental absorption peak,
$\lambda_\mathrm{max}$.

It is interesting to compare the absolute intensities of the spectra
obtained from different solvation models: the resulting trend
$I_\mathrm{MSM}>I_\mathrm{CSM}>I_\mathrm{deh}$ can be ascribed to the
larger delocalization of molecular orbitals \emph{in vacuo} than in
solution, which implies a smaller excitation oscillator
strength. Molecular orbitals are more delocalized in CSM than in MSM,
because the latter model explicitly accounts for Pauli repulsion from
the solvent molecules, whereas the former does not (at least in our
implementation).

\begin{figure}[ht!]
  \begin{center}
    \includegraphics[width=0.4\textwidth]{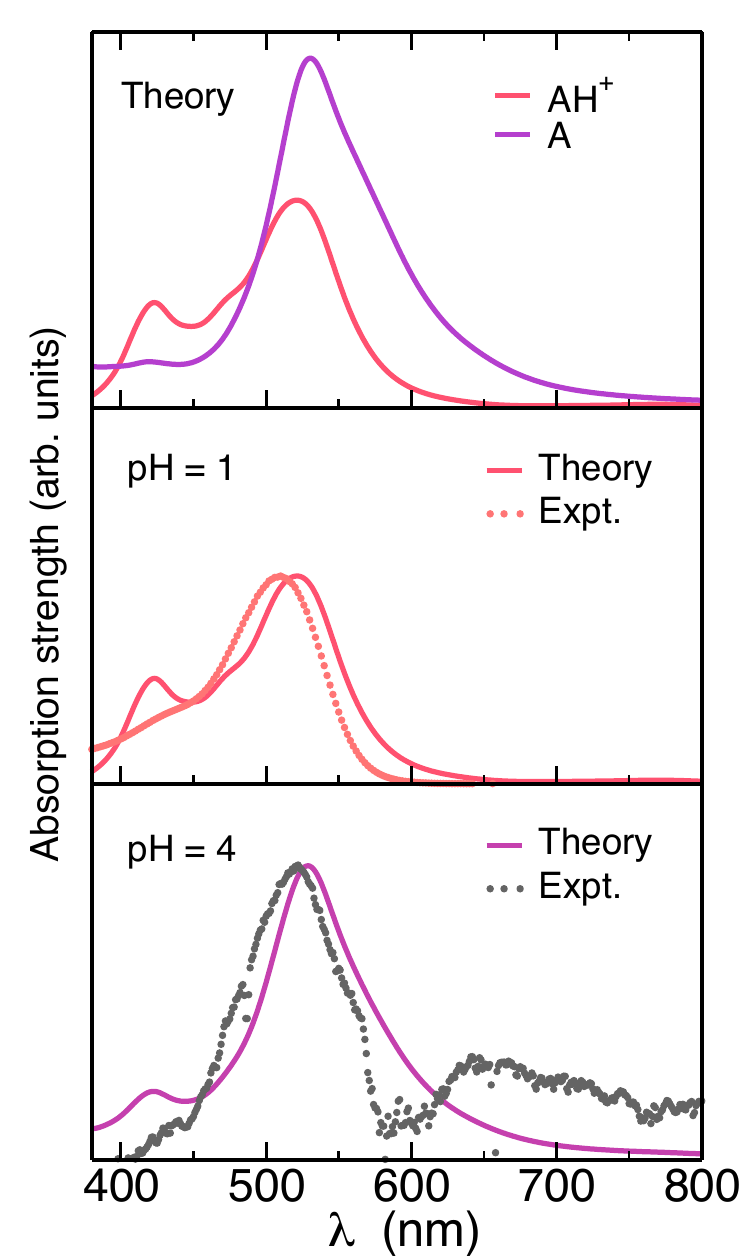}
  \end{center}
  \caption{Upper panel:~Comparison of the time-averaged theoretical
    absorption spectra for A and AH$^+$ computed with the
    multimodel approach and using the B3LYP XC functional. Middle
    panel:~Comparison of the theoretical and
    experimental\cite{Robbins:2015} spectra for AH$^+$ cation at pH=1.
    Lower panel:~Comparison of the theoretical and
    experimental\cite{Robbins:2015} spectra for neutral A base at pH=4.  The
    color code corresponds to the color resulting from the tristimulus
    colorimetry analysis, except for the experimental spectrum at
    pH=4, which is shown by grey full dots since the solution is
    almost transparent in this case. The relative intensities on all
    theoretical spectra are respected, while the intensities of the
    experimental spectra were adjusted to the theoretical ones for the
    first absorption peak.}
  \label{fig:expt}
\end{figure}

Checking the convergence of our time-averaged absorption spectra
requires some care. In Figure~\ref{fig:convergence} we display the
absorption spectra of cyanin computed in MSM and CSM using the PBE
functional, and averaged over an increasing number of consecutive
frames, evenly spaced by 0.5~ps. We see that up to $\approx 60$
independent frames are needed to achieve a reasonable
convergence. When not otherwise specified, the spectra presented in
the text are obtained by averaging over 100 configurations. The
convergence of the time-averaged spectrum in CSM using B3LYP XC
functional follows the same trend.

\section{Predicting experimental spectra} \label{sec:expt}

Figure~\ref{fig:expt}~(upper panel) shows the spectra predicted by our
multimodel approach for the solvated cyanin, in both the flavylium cation
(pink), and quinonoidal neutral base (purple) charge states, using the
B3LYP XC functional. At very low pH (<3), the flavylium cation, is the
only stable species present in solution, and our theoretical predictions
for its spectrum can thus be directly compared with experiment
(Figure~\ref{fig:expt}, middle panel). The agreement with the
experiments is very good as concerns both the overall shape of the
spectrum, including the position of the maximum-absorption peak
($\lambda_\mathrm{max}^\mathrm{theor} = 520$~nm,
$\Delta\lambda_\mathrm{max}\approx 12$~nm), and the hue of its
simulated color. Although our simulations predict a secondary
absorption peak at shorter wave-length
($\lambda\approx 425~\mathrm{nm}$) that is not observed
experimentally, a shoulder is clearly visible in the experimental
spectrum at this wave-length.\cite{Robbins:2015}

As the acidity of the solution decreases, different molecular species
coexist: \emph{i)} deprotonation of the phenyl ring can take place
giving rise to the quinonoidal base; \emph{ii)} hydration of the
quinonoidal base generates several optically inactive, colorless
species (e.g. chalcones).\cite{Brouillard:1988} The pH-dependent molar
fractions of these species have been experimentally determined in
Ref. \citenum{Leydet:2012}. At $\mathrm{pH}=4$ the relative
concentrations of the various species involved in the chemical
equilibrium were found out to be:
$\mathrm{\left[ AH^+ \right] = 3\%}$,
$\mathrm{\left [ A \right] = 5\%}$, and
$\mathrm{\left [ CL \right] = 92 \%}$, where [CL] indicates the molar
fraction of the ensemble of the colorless components.  The high
percentage of colorless components bleaches the color of the overall
solution.  A residual optical activity, however, still persists, and
our predictions for the optical properties of the quinonoidal base can
be validated by comparing the weighted sum of the spectra computed for
the flavylium cation and quinonoidal base with the measured
absorptivity. Our results are shown in Figure~\ref{fig:expt}~(lower
panel), which displays once again a very good agreement with
experiments:\cite{Robbins:2015}
$\lambda_\mathrm{max}^\mathrm{theor} = 529$~nm,
$\Delta\lambda_\mathrm{max}\approx 8$~nm (actually even better than at
very low pH, probably as a result of a fortuitous cancellation of
errors).

\section{Conclusions} \label{sec:conclusions}

In this paper we have introduced a new approach to simulate the
optical properties of complex dyes in solution in the visible range,
based on a multimodel protocol to account for the effects of the
solute-solvent interactions (such as thermal vibrations and dielectric
screening) on the absorption spectra of molecular dyes. Our
  methodology builds on previous attempts to simulate thermal
  fluctuation effects using classical\cite{Barone:2010,DeMitri:2013,Cacelli:2016}
  or \emph{ab initio}\cite{Malcioglu:2011b} molecular dynamics and
  TDDFT to compute the optical spectra for individual molecular
  configurations thus generated, and improves upon them by avoiding to
  use classical force fields, one one hand, and by optimally
  combining explicit and implicit solvation models on the other hand. The
  validation of the newly introduced protocol, performed on
  cyanidin-3-glucoside in its neutral (quinonoidal base) charge state
  at room temperature, sheds light onto the validity of implicit
  (continuum) solvation models of the structural and spectroscopic
  properties of molecular dyes in solution, and results in a
  remarkable agreement with the experiments in the strong-to-weak
  acidity range.

Our developments open the way to the simulation of the optical
properties of complex solvated dyes in more realistic conditions, and
with a higher accuracy than it has been possible so far. Our approach,
which is based on extensive sampling of the optical properties of the
solute corresponding to different molecular configurations occurring
at equilibrium, is being further enhanced by accounting for
low-frequency conformational fluctuations occurring over widely varying
time scales, \cite{Rosa:2016} by developing new ``learn-on-the-fly''
techniques to save on the time-consuming TDDFT calculations over
molecular dynamics trajectories,\cite{Micciarelli:2016} and by
combining it with a QM/MM approach to solute-solvent interactions.

\begin{acknowledgement}
  We are grateful to Rebecca Robbins and Tom Collins for illuminating
  discussions and for providing us with their unpublished data of
  Ref. \citenum{Robbins:2015}. We also thank Xiaochuan Ge and Oliviero
  Andreussi for cooperating with us during some of the preliminary
  phases of the present work, and Carlo Cavazzoni and Fabio
  Affinito for technical support at CINECA. This work was partially
  supported by {\it Mars Chocolate North America LLC} and by the
  European Union through the \textsc{MaX} Centre of Excellence (Grant
  No. 676598). Computer resources were partially provided by the CINECA
  Supercomputing Center (Italy) at their Fermi BG/Q Tier-0 and Galileo Tier-1
  machines through grants PRACE No. 2013081532 ({\it Chromatology}),
  IscrC\_MULAN, IscrC\_SCCS-H, and IscrC\_SCCS-H2.
\end{acknowledgement}

\providecommand{\latin}[1]{#1}
\providecommand*\mcitethebibliography{\thebibliography}
\csname @ifundefined\endcsname{endmcitethebibliography}
  {\let\endmcitethebibliography\endthebibliography}{}

%\newpage\begin{tocentry}
%  \includegraphics[width=\textwidth]{TOC_figure.pdf}
%\end{tocentry}

\end{document}